\documentclass[conference]{IEEEtran}
\IEEEoverridecommandlockouts
\usepackage[subtle,tracking=normal,charwidths=tight]{savetrees}
\usepackage{anyfontsize}
\usepackage{cite}
\usepackage{amsmath,amssymb,amsfonts}
\usepackage{algorithmic}
\usepackage{graphicx}
\usepackage{textcomp}
\usepackage{xcolor}
\usepackage{siunitx}
\usepackage{pgfplots,tikz}
\pgfplotsset{compat=1.16}
\def\BibTeX{{\rm B\kern-.05em{\sc i\kern-.025em b}\kern-.08em
    T\kern-.1667em\lower.7ex\hbox{E}\kern-.125emX}}
\newcommand{\vej}{\mathbf{e}_{j}}
\newcommand{\veOne}{\mathbf{e}_{1}}
\newcommand{\vejHead}{\mathbf{e}_{\rm{ h},j}}
\newcommand{\veOneHead}{\mathbf{e}_{\rm{ h},1}}
\newcommand{\vn}{\mathbf{n}\left(l\right)}
\newcommand{\vnHermit}{\mathbf{n}^{H}\left(l\right)}
\newcommand{\vx}{\mathbf{x}\left(l\right)}
\newcommand{\vxHermit}{\mathbf{x}^{H}\left(l\right)}
\newcommand{\vy}{\mathbf{y}\left(l\right)}
\newcommand{\vyHermit}{\mathbf{y}^{H}\left(l\right)}
\newcommand{\vg}{\mathbf{g}\left(l\right)}
\newcommand{\vgHermit}{\mathbf{g}^{H}\left(l\right)}
\newcommand{\vgHead}{\mathbf{g}_{\rm h}\left(l\right)}

\newcommand{\Rn}{\boldsymbol{\Phi}_{\rm n}\left(l\right)}
\newcommand{\Rx}{\boldsymbol{\Phi}_{\rm x}\left(l\right)}
\newcommand{\Ry}{\boldsymbol{\Phi}_{\rm y}\left(l\right)}
\newcommand{\Rnhat}{\hat{\boldsymbol{\Phi}}_{\rm n}\left(l\right)}

\newcommand{\Ryhat}{\hat{\boldsymbol{\Phi}}_{\rm y}\left(l\right)}
\newcommand{\RywhiteHat}{\hat{\boldsymbol{\Phi}}_{\rm y}^{\rm (w)}\left(l\right)}
\newcommand{\RywhiteHeadHat}{\hat{\boldsymbol{\Phi}}_{\rm y,h}^{\rm (w)}\left(l\right)}
\newcommand{\RnHead}{\boldsymbol{\Phi}_{\rm n,h}\left(l\right)}
\newcommand{\RxHead}{\boldsymbol{\Phi}_{\rm x,h}\left(l\right)}
\newcommand{\RyHead}{\boldsymbol{\Phi}_{\rm y,h}\left(l\right)}
\newcommand{\RnHeadHat}{\hat{\boldsymbol{\Phi}}_{\rm n,h}\left(l\right)}

\newcommand{\RyHeadHat}{\hat{\boldsymbol{\Phi}}_{\rm y,h}\left(l\right)}

\begin{document}

\title{Comparison of Binaural RTF-Vector-Based \\ Direction of Arrival Estimation Methods \\ Exploiting an External Microphone
\thanks{Funded by the Deutsche Forschungsgemeinschaft (DFG, German \mbox{Research} Foundation) - Project ID 352015383 (SFB 1330 B2) and Project ID 390895286 (EXC 2177/1).}
}

\author{\IEEEauthorblockN{Daniel Fejgin and Simon Doclo}
\IEEEauthorblockA{Department of Medical Physics and Acoustics and Cluster of Excellence Hearing4All, University of Oldenburg, Germany}
\{daniel.fejgin,simon.doclo\}@uol.de}

\maketitle

\begin{abstract}
In this paper we consider a binaural hearing aid setup, where in addition to the head-mounted microphones an external microphone is available. For this setup, we investigate the performance of several relative transfer function (RTF) vector \mbox{estimation} methods to estimate the direction of arrival (DOA) of the target speaker in a noisy and reverberant acoustic environment. More in particular, we consider the state-of-the-art covariance whitening (CW) and covariance subtraction (CS) methods, either incorporating the external microphone or not, and the recently proposed spatial coherence (SC) method, requiring the external microphone. To estimate the DOA from the estimated RTF vector, we propose to minimize the frequency-averaged Hermitian angle between the estimated head-mounted RTF vector and a database of prototype head-mounted RTF vectors. Experimental results with stationary and moving speech sources in a reverberant environment with diffuse-like noise show that the SC method outperforms the CS method and yields a similar DOA estimation accuracy as the CW method at a lower computational complexity.
\end{abstract}


\section{Introduction}
For binaural hearing aid (HA) applications, estimating the direction of arrival (DOA) of the target speaker in a noisy and \mbox{reverberant} acoustic environment is important, e.g., to steer a beamformer towards this speaker \cite{Marquardt2016}. Several methods have been proposed for binaural DOA estimation, e.g., based on inter\-aural time and level differences \cite{Raspaud2010,May2012}, generalized cross-correlation (GCC) features \cite{Kayser2014,Marquardt2017,Ma2018,Varzandeh2020}, or relative transfer function (RTF) vectors \cite{Braun2015}. For a binaural HA setup incorporating an external microphone, an RTF-vector-based DOA estimation method was proposed in \cite{Farmani2018}, where it was however assumed that the external microphone was worn by the target speaker, such that the external microphone signal almost does not capture any noise or reverberation.

To estimate the RTF vector of the target speaker from noisy microphone signals, several methods have been proposed in the literature \cite{Cohen2004,Markovich2009,Krueger2011,Varzandeh2017,Golan2018}, where the most popular methods are based on covariance subtraction (CS) or covariance whitening (CW). These methods require an estimate of the covariance \mbox{matrix} of the noisy microphone signals (e.g., estimated during speech-plus-noise time-frequency (TF) bins) and the noise covariance matrix (e.g., estimated during noise-only TF bins). In should be realized that due to the involved eigenvalue decomposition the computational complexity of the CW method is in general high, which is especially relevant for an online implementation. Exploiting the availability of an external microphone, in \cite{Goessling2018,Goessling2020} the spatial coherence (SC) method was proposed to estimate the (head-mounted) RTF vectors. The SC method relies on the assumption that the coherence between the noise component in the external microphone signal and the noise components in the head-mounted microphone signals is low. This assumption holds quite well, for example, when the distance between the external microphone and the head-mounted microphones is large enough and the noise field is diffuse-like. In comparison to the CS and CW methods an additional advantage is the fact that no estimate of the noise covariance matrix is required. 

For a binaural hearing aid setup with an external microphone that is not worn by the target speaker, in this paper we analyze the performance of several RTF-vector-based DOA estimation methods, more in particular, CS and CW (either incorporating the external microphone or not) and SC (\mbox{incorporating} the external microphone). Instead of using a statistical classifier or a neural network to estimate the DOA from the estimated RTF vectors \cite{Li2016,Hammer2020}, we follow an approach similar to \cite{Marquardt2017,Braun2015}, where the estimated head-mounted RTF vectors are compared to a database of anechoic prototype RTF vectors for several directions. However, instead of using a least-squares-based cost function, we propose to use a cost function based on the Hermitian angle.  Experimental results using recorded signals in a reverberant environment with diffuse-like noise show that the SC method outperforms the CS method and yields a similar DOA estimation accuracy as the more computational complex CW method, both for a static as well as for a moving target speaker and for several positions of the external microphone.

\section{Signal Model}
\label{seq:sigMod}
We consider a binaural hearing aid setup consisting of $M$ head-mounted microphones and one external microphone, which is spatially separated from the head-mounted microphones, thus, $M + 1$ microphones in total. We consider a single speech source at DOA $\theta_{\rm s}$ (in the azimuthal plane) in a noisy and reverberant acoustic environment, see Fig. \ref{fig:setup}. The $m$-th microphone signal can be written in the short-time Fourier transform (STFT) domain as
\begin{align} 
	\label{eq:signalModel_micComponent}
	Y_{m}\left(k,l\right) = X_{m}\left(k,l\right) + N_{m}\left(k,l\right)\,, \quad m \in \left\{1,\dots,M+1\right\}\,,
\end{align}
where the speech and noise components at the $k$-th frequency bin $\left(k \in \left\{1,\dots,K\right\}\right)$ and the $l$-th frame $\left(l \in \left\{1,\dots,L\right\}\right)$ are denoted by $X_{m}\left(k,l\right)$ and $N_{m}\left(k,l\right)$, respectively. Since all frequency bins are assumed to be independent and are hence treated independently, we will omit the index $k$ in the remainder of the paper where possible. Stacking the $M+1$ microphone signals in a vector $\vy =\left[Y_{1}\left(l\right),\,\dots,\,Y_{M+1}\left(l\right)\right]^{T}$, where $\left(\cdot\right)^{T}$ denotes transposition, and defining $\vx$ and $\vn$ similarly as $\vy$, the vector $\vy$ can be written as
\begin{equation}
	\label{eq:signalModel}
	\vy = \vx + \vn \in \mathbb{C}^{M+1}\,.
\end{equation}
Assuming that the multiplicative transfer function approximation \cite{Avargel2007} holds, the speech vector $\vx$ can be written as
\begin{equation}
	\label{eq:speechModel}
	\vx = \vg X_{1}\left(l\right)\,,
\end{equation}
where the $\left(M+1\right)$-dimensional \textit{extended} RTF vector
\begin{equation}
	\label{eq:defRTFext}
	\vg = \left[1,\,G_{2}\left(l\right),\,\dots,\,G_{M+1}\left(l\right)\right]^{T}
\end{equation}
contains the reverberant RTFs of the speech source between all microphones (including the external microphone) and the reference microphone, for which we have used the first microphone without loss of generality. The $M$-dimensional \textit{head-mounted} RTF vector $\vgHead$ corresponding to the head-mounted microphones can be extracted from $\vg$ in \eqref{eq:defRTFext} as
\begin{equation}
	\label{eq:defRTFHead}
	\vgHead = \mathbf{E}\vg\,, \quad \mathbf{E} = \left[\mathbf{I}_{M \times M},\,\mathbf{0}_{M}\right]\,,
\end{equation}
where $\mathbf{I}_{M \times M}$ is the $M \times M$-dimensional identity matrix and $\mathbf{0}_{M}$ is the $M$-dimensional zero vector. Since it can be assumed that the relative positions of the head-mounted microphones are fixed (ignoring small movements of the hearing aids due to head movements) whereas the external microphone can be located at an arbitrary position, it should be realized that although the extended RTF vector $\vg$ encodes the DOA $\theta_{\rm s}$, it depends on the (unknown) position of the external microphone, whereas the head-mounted RTF vector $\vgHead$ encodes the DOA $\theta_{\rm s}$ and obviously does not depend on the position of the external microphone. Hence, for DOA estimation, we will only consider the head-mounted RTF vector.

The $\left(M+1\right) \times \left(M+1\right)$-dimensional speech and noise covariance matrices are defined as
\begin{align}
	\Rx &= \mathcal{E}\{\vx\vxHermit\} = \vg\vgHermit\Phi_{X_{1}}\left(l\right),\,\label{eq:SOS_x}\\
	\Rn &= \mathcal{E}\{\vn\vnHermit\}\,,\label{eq:SOS_n}
\end{align}
where $\Phi_{X_{1}}\left(l\right) =\mathcal{E}\left\{\lvert X_{1}\left(l\right)\rvert^{2}\right\}$ denotes the power spectral density of the speech component in the reference microphone signal, and the operators $\left(\cdot\right)^{H}$ and $\mathcal{E}\left\{\cdot\right\}$ denote complex transposition and expectation, respectively. Assuming un\-correlated speech and noise components, the covariance matrix of the noisy microphone signals $\Ry$ can be written as
\begin{equation}
	\label{eq:SOS}
	\Ry = \mathcal{E}\{\vy\vyHermit\} = \Rx + \Rn\,.
\end{equation}
The $M \times M$-dimensional covariance matrices corresponding to the head-mounted microphones can be extracted from \eqref{eq:SOS_x} - \eqref{eq:SOS} as
\begin{align}
	\RxHead &= \mathbf{E}\Rx\mathbf{E}^{T}\,, \quad \RnHead = \mathbf{E}\Rn\mathbf{E}^{T}\,,\\  
	\RyHead &= \mathbf{E}\Ry\mathbf{E}^{T} = \RxHead + \RnHead\,.
\end{align}

\section{RTF Vector Estimation}
\label{seq:rtf}
In this section we discuss several RTF vector estimation \mbox{methods}. In Sections \ref{subsec:CS} and \ref{subsec:CW} we review the state-of-the-art covariance subtraction (CS) and covariance whitening (CW) methods \cite{Cohen2004,Markovich2009,Golan2018}, which are general methods that can be used to estimate the extended RTF vector (using all microphones) or the head-mounted RTF vector (using only the head-mounted microphones). In Section \ref{subsec:SC} we discuss the recently proposed spatial coherence method \cite{Goessling2018,Goessling2020}, which requires the availability of an external microphone to estimate the head-mounted RTF vector.
\subsection{Covariance Subtraction (CS)}
\label{subsec:CS}
Using \eqref{eq:SOS_x} and \eqref{eq:SOS}, the extended RTF vector $\vg$ can be obtained from any column of the rank-1 speech covariance matrix $\Rx$ with appropriate normalization \cite{Cohen2004,Golan2018}, i.e.,
\begin{equation}
	\label{eq:CSPerf}
	\vg = \frac{\Rx\vej}{\veOne^{T}\Rx\vej} = \frac{\left(\Ry-\Rn\right)\vej}{\veOne^{T}\left(\Ry-\Rn\right)\vej}\,,
\end{equation}
where $\vej = \left[0,\,\dotsc,\,1,0\dotsc,\,0\right]^{T}$ is an $\left(M+1\right)$-dimensional vector with zeros except the $j$-th element. In practice, estimates of the noisy covariance matrix $\Ryhat$ and the noise covariance matrix $\Rnhat$ are used (e.g., obtained via recursive smoothing during speech-plus-noise and noise-only TF bins), yielding the CS estimate of the extended RTF vector
\begin{equation}
	\label{eq:CS_extended}
	\hat{\mathbf{g}}^{\left(\rm CS\right)}\left(l\right) = \frac{\left(\Ryhat - \Rnhat\right)\vej}{\veOne^{T}\left(\Ryhat - \Rnhat\right)\vej}\,.
\end{equation}
Similarly, when using the covariance matrices corresponding to the head-mounted microphones (i.e., not exploiting the external microphone), the CS estimate of the head-mounted RTF vector is given by
\begin{equation}
	\label{eq:CS_head}
	\boxed{\hat{\mathbf{g}}_{\rm h}^{\left(\rm CS\right)}\left(l\right) = \frac{\left(\RyHeadHat - \RnHeadHat\right)\vejHead}{\veOneHead^{T}\left(\RyHeadHat - \RnHeadHat\right)\vejHead}}
\end{equation}
where $\vejHead = \left[0,\,\dotsc,\,1,0\dotsc,\,0\right]^{T}$ is an $M$-dimensional vector with zeros except the $j$-th element. It can be easily shown that
\begin{equation}
	\hat{\mathbf{g}}_{\rm h}^{\left(\rm CS\right)}\left(l\right) = \mathbf{E}\hat{\mathbf{g}}^{\left(\rm CS\right)}\left(l\right)\,,
\end{equation}
such that this estimate does not depend on the position of the external microphone. Hence, in the experiments in Section \ref{seq:exp} we will only consider one version of the CS method (without the external microphone).
\subsection{Covariance Whitening (CW)}
\label{subsec:CW}
Instead of subtracting $\Rnhat$ from $\Ryhat$, the CW method first prewhitens the estimated noisy covariance matrix with a square-root decomposition (e.g., Cholesky decomposition) of the estimated noise covariance matrix \cite{Markovich2009,Golan2018}, i.e.,
\begin{equation}
	\Rnhat = \hat{\mathbf{L}}_{\rm n}\left(l\right)\hat{\mathbf{L}}_{\rm n}^{H}\left(l\right)\,, \quad \RywhiteHat =\hat{\mathbf{L}}_{\rm n}^{-1}\left(l\right)\Ryhat\hat{\mathbf{L}}_{\rm n}^{-H}\left(l\right)\,.\label{eq:CW_preWork}
\end{equation} 
The CW estimate of the extended RTF vector is then obtained as the normalized de-whitened principal eigenvector of the pre-whitened noisy covariance matrix, i.e.,
\begin{equation}
	\label{eq:CW_extended}
	\boxed{\hat{\mathbf{g}}^{\left(\rm CW\right)}\left(l\right) =\frac{\hat{\mathbf{L}}_{\rm n}\left(l\right)\mathcal{P}\left\{\RywhiteHat\right\}}{\veOne^{T}\hat{\mathbf{L}}_{\rm n}\left(l\right)\mathcal{P}\left\{\RywhiteHat\right\}}}
\end{equation}
where $\mathcal{P}\left\{\cdot\right\}$ denotes the principal eigenvector of a matrix.
 
Similarly, when using the covariance matrices corresponding to the head-mounted microphones (i.e., not exploiting the external microphone), the CW estimate of the head-mounted RTF vector is given by 
\begin{equation}
	\label{eq:CW_head}
	\boxed{\hat{\mathbf{g}}_{\rm h}^{\left(\rm CW\right)}\left(l\right) =\frac{\hat{\mathbf{L}}_{\rm n,h}\left(l\right)\mathcal{P}\left\{\RywhiteHeadHat\right\}}{\veOneHead^{T}\hat{\mathbf{L}}_{\rm n,h}\left(l\right)\mathcal{P}\left\{\RywhiteHeadHat\right\}}}
\end{equation}
with
\begin{equation}
	\RnHeadHat = \hat{\mathbf{L}}_{\rm n,h}\left(l\right)\hat{\mathbf{L}}_{\rm n,h}^{H}\left(l\right)\,, \quad \RywhiteHeadHat =\hat{\mathbf{L}}_{\rm n,h}^{-1}\left(l\right)\RyHeadHat\hat{\mathbf{L}}_{\rm n,h}^{-H}\left(l\right)\,.\label{eq:CW_preWork2}
\end{equation}
Since contrary to the CS method 
\begin{equation}
	\hat{\mathbf{g}}_{\rm h}^{\left(\rm CW\right)}\left(l\right) \neq \mathbf{E}\hat{\mathbf{g}}^{\left(\rm CW\right)}\left(l\right)
\end{equation}
in the experiments in Section \ref{seq:exp} we will consider two versions of the CW method, either exploiting the external microphone or not. Due to the required square-root decomposition in \eqref{eq:CW_preWork} or \eqref{eq:CW_preWork2} and the eigenvalue decomposition in \eqref{eq:CW_extended} or \eqref{eq:CW_head}, the computational complexity for the CW method is larger than for the CS method.
\subsection{Spatial Coherence (SC)}
\label{subsec:SC}
The SC method \cite{Goessling2018,Goessling2020} requires an external microphone and assumes a low coherence between the noise component in the external microphone signal and the noise components in the head-mounted microphone signals, i.e.,
\begin{equation}
	\label{eq:SC_mainAssumption}
	\mathcal{E}\left\{N_{i}\left(l\right)N_{M+1}^{\ast}\left(l\right)\right\} \approx 0\,, \quad i \in \left\{1,\dots,M\right\}\,,
\end{equation}
As shown in \cite{Goessling2018,Goessling2020}, this assumption holds quite well for a diffuse-like noise field (e.g., multi-talker babble noise) when the distance between the external microphone and the head-mounted microphones is large enough. Using \eqref{eq:SC_mainAssumption}, it can be easily shown that
\begin{equation}
	\label{eq:SC_assumption2}
	\mathcal{E}\left\{Y_{i}\left(l\right)Y_{M+1}^{\ast}\left(l\right)\right\} = \mathcal{E}\left\{X_{i}\left(l\right)X_{M+1}^{\ast}\left(l\right)\right\}\,, ~~ i \in \left\{1,\dots,M\right\}\,,
\end{equation}
such that, using \eqref{eq:SOS_x}, the head-mounted RTF vector can be estimated from the $\left(M+1\right)$-th column of the estimated noisy covariance matrix $\Ryhat$ as
\begin{equation}
	\label{eq:SC}
	\boxed{\hat{\mathbf{g}}_{\rm h}^{\left(\rm SC\right)}\left(l\right) = \mathbf{E}\frac{\Ryhat\mathbf{e}_{M+1}}{\veOne^{T}\Ryhat\mathbf{e}_{M+1}}}
\end{equation}
The SC method has a similar computational complexity as the CS method and a lower complexity as the CW method, but contrary to the CS and CW method does not require an estimate of the noise covariance matrix $\Rnhat$.

\section{DoA Estimation}
\label{seq:doa}
To estimate the possibly time-varying DOA $\theta_{\rm s}\!\left(l\right)$ of the target speaker from the estimated head-mounted RTF vector $\hat{\mathbf{g}}_{\rm h}\!\left(k,l\right)$, different approaches have been proposed\footnote{As already mentioned, since the estimated extended RTF vector $\hat{\mathbf{g}}\left(k,l\right)$ depends on the (unknown) position of the external microphone, it cannot be straightforwardly used for DOA estimation.}. Instead of using a statistical classifier or a neural network as in \cite{Li2016,Hammer2020}, in \cite{Marquardt2017,Braun2015} it has been proposed to simply compare the estimated head-mounted RTF vector with a database of anechoic prototype head-mounted RTF vectors $\bar{\mathbf{g}}_{\rm h}\left(k,\theta_{i}\right)$ for different discrete directions $\theta_{i}\,,~i=1,\dots,I$. These prototype head-mounted RTF vectors can either be obtained using, e.g., a spherical diffraction model \cite{Duda1998}, or measured using the same microphone array configuration as used during the actual source localization. Whereas the cost functions in \cite{Marquardt2017,Braun2015} use the (squared) norm between the (normalized) estimated and prototype head-mounted RTF vectors, in this paper we propose to use the so-called Hermitian angle \cite{Varzandeh2017} between the estimated and prototype head-mounted RTF vectors, i.e.,
\begin{equation}
	\label{eq:hermitAngle_narrowband}
	d\left(k,l,\theta_{i}\right) = \arccos\left(\frac{\lvert\bar{\mathbf{g}}_{\rm h}^{H}\left(k,\theta_{i}\right)\hat{\mathbf{g}}_{\rm h}\left(k,l\right)\rvert}
	{\lVert\bar{\mathbf{g}}_{\rm h}\left(k,\theta_{i}\right)\rVert_{2}\,\lVert\hat{\mathbf{g}}_{\rm h}\left(k,l\right)\rVert_{2}}\right)\,,
\end{equation} 
since this resulted in a better DOA estimation accuracy. The DOA of the target speaker is then estimated as the direction for which the frequency-averaged cost function in \eqref{eq:hermitAngle_narrowband} is minimal, i.e.,
\begin{equation}
	\label{eq:doaEst}
	\hat{\theta}_{\rm s}\left(l\right) = \underset{\theta_{i}}{\rm{argmin}}~J\left(l,\theta_{i}\right) =  \underset{\theta_{i}}{\rm{argmin}}~\frac{1}{K-1}\sum_{k=2}^{K}d\left(k,l,\theta_{i}\right)\,.
\end{equation}

\vspace{-0.35em}
\section{Experimental Results}
\label{seq:exp}
In this section we compare the DOA estimation accuracy using four different RTF vector estimates: 
\begin{itemize}
	\item The CS-based estimate $\hat{\mathbf{g}}_{\rm h}^{\left(\rm CS\right)}\left(l\right)$ in \eqref{eq:CS_head} using only the head-mounted microphones. It should be noted that this is similar to the binaural DOA estimation method presented in \cite{Braun2015}.
	\item The CW-based estimates $\mathbf{E}\hat{\mathbf{g}}^{\left(\rm CW\right)}\left(l\right)$ based on \eqref{eq:CW_extended}, using all microphones, and $\hat{\mathbf{g}}_{\rm h}^{\left(\rm CW\right)}\left(l\right)$ in \eqref{eq:CW_head} using only the head-mounted microphones.
	\item The SC-based estimate $\hat{\mathbf{g}}_{\rm h}^{\left(\rm SC\right)}\left(l\right)$ in \eqref{eq:SC} using all microphones.

\end{itemize}
The experimental setup and implementation details are described in Section \ref{sec:exp_Details}. Experimental results for a static and a moving speaker in a reverberant environment with diffuse-like noise are presented in Section \ref{sec:exp_res}.

\subsection{Experimental setup and implementation details}
\label{sec:exp_Details}
\begin{figure}[t]
		\centering
		\includegraphics[width=0.2\textwidth,trim={5.75cm 5cm 9.5cm 5cm},clip]{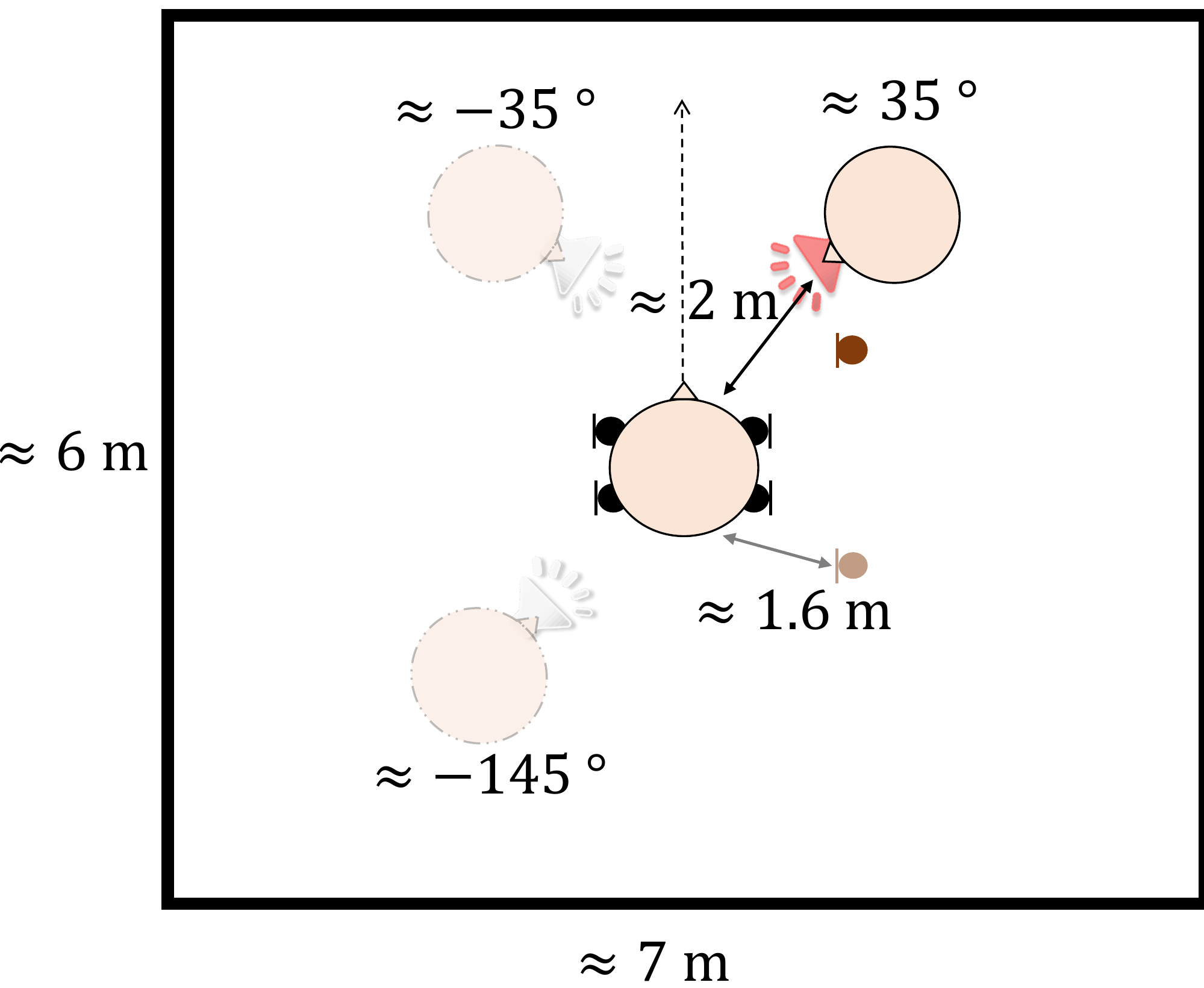}
		\includegraphics[width=0.1725\textwidth,trim={8.5cm 5cm 9.5cm 5cm},clip]{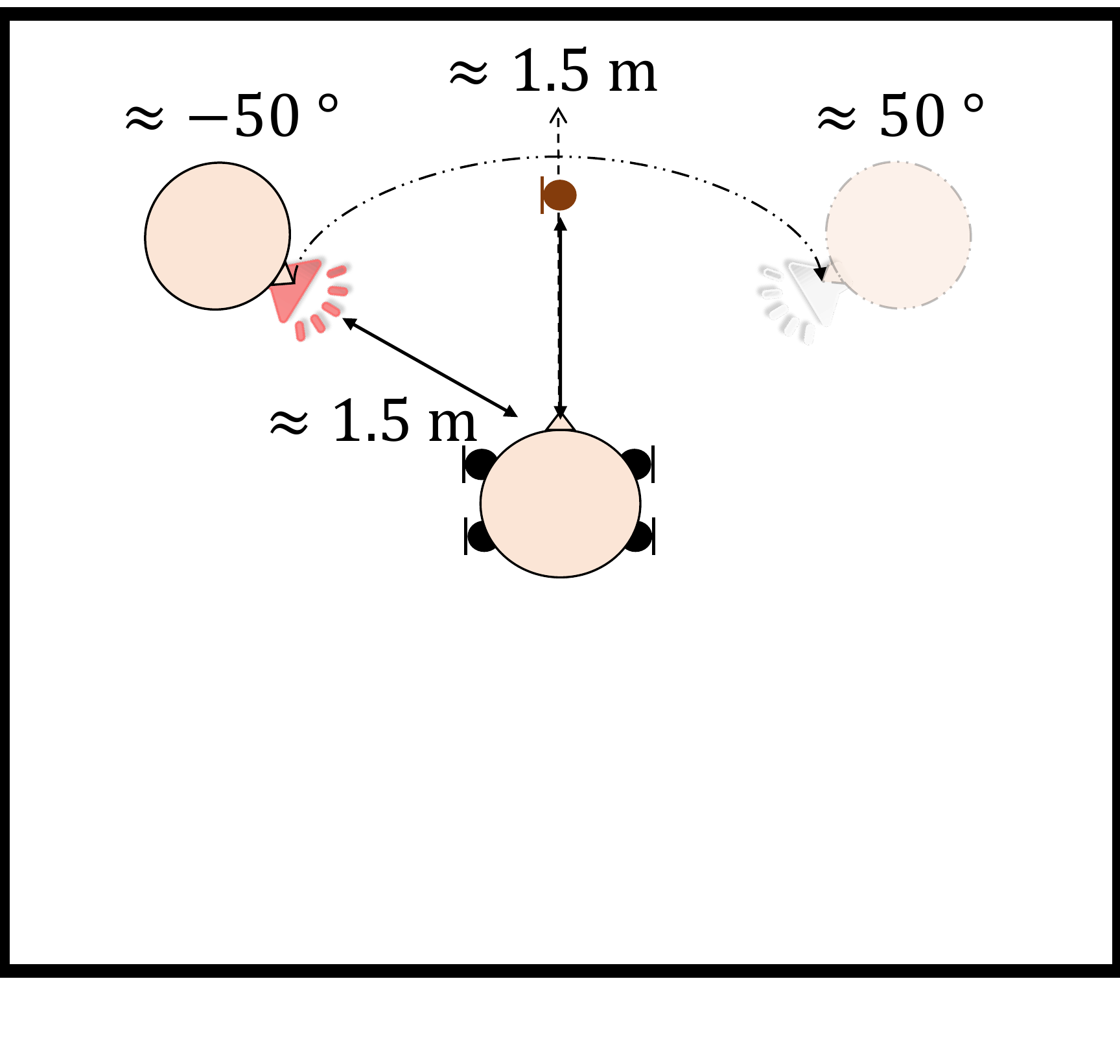}
		\vspace{-3mm}
		\caption{Experimental setup for stationary speaker scenario (left) and moving speaker scenario (right). The external microphone is depicted in brown whereas the head-mounted microphones are depicted in black.}
		\vskip-0.34cm
	\label{fig:setup}
\end{figure}
For the experiments we used recordings in a laboratory at the University of Oldenburg with dimensions about $\SI[product-units=brackets]{7x6x2.7}{\cubic\metre}$, where the reverberation time can be easily changed by closing and opening absorber panels mounted to the walls and ceiling. Fig. \ref{fig:setup} depicts the experimental setup, where a dummy head with bi\-naural hearing aids ($M=4$ microphones) is located approximately in the center of the laboratory. The external microphone is not restricted to be close to the target speaker. We consider two scenarios, either with a stationary speaker or with a moving speaker. For both scenarios, the speech and noise components were recorded se\-parately. Diffuse-like noise was generated with four loud\-speakers facing the corners of the laboratory, playing back different multi-talker recordings. The signal-to-noise ratio (SNR) was set as the ratio of the average broadband speech power to broadband noise power in the front microphones of both hearing aids.

For the stationary speaker scenario, three different positions of the speech source and two different positions of the external microphone are considered (see Fig. \ref{fig:setup}). The speech source is located at approximately $\SI{2}{\metre}$ from the dummy head at either $\ang{-145}$, $\ang{-35}$, or $\ang{35}$. The external microphone is located at approximately $\SI{1.6}{\metre}$ from the dummy head at either $\ang{45}$ or $\ang{130}$. The speech source is constantly active and comprises English sentences (duration: $\SI{30}{\second}$).

For the moving speaker scenario, a male speaker moves from approximately $\ang{-50}$ to $\ang{50}$ at a distance of about $\SI{1.5}{\metre}$ from the dummy head (see Fig. \ref{fig:setup}). The external microphone is located at approximately $\SI{1.5}{\metre}$ in front of the dummy head. The speaker is constantly active (duration: $\SI{25}{\second}$).

The microphone signals are recorded at a sampling frequency $f_{\rm s} =\SI{16}{\kilo\hertz}$ and processed in the STFT-domain using a $32 \: \si{\milli\second}$ square-root Hann window with $\SI{50}{\percent}$ overlap. The noisy and noise covariance matrices are recursively estimated during detected speech-plus-noise and noise-only TF-bins, respectively, as in \eqref{eq:SOS_RyUpdate} and \eqref{eq:SOS_RnUpdate} using smoothing factors $\alpha_{\rm y}$ and $\alpha_{\rm n}$ corresponding to time constants of $\SI{250}{\milli\second}$ for $\hat{\boldsymbol{\Phi}}_{\rm y}\left(k,l\right)$ and $\SI{500}{\milli\second}$ for $\hat{\boldsymbol{\Phi}}_{\rm n}\left(k,l\right)$ for the stationary speaker scenario and using smoothing factors corresponding to time constants of $\SI{150}{\milli\second}$ for $\hat{\boldsymbol{\Phi}}_{\rm y}\left(k,l\right)$ and $\SI{500}{\milli\second}$ for $\hat{\boldsymbol{\Phi}}_{\rm n}\left(k,l\right)$ for the moving speaker scenario. 
\begin{align}
	\hat{\boldsymbol{\Phi}}_{\rm y}\left(k,l\right) &= \alpha_{\rm y}\hat{\boldsymbol{\Phi}}_{\rm y}\left(k,l-1\right) + \mathbf{y}\left(k,l\right)\mathbf{y}^{H}\left(k,l\right)\label{eq:SOS_RyUpdate}\\
	\hat{\boldsymbol{\Phi}}_{\rm n}\left(k,l\right) &= \alpha_{\rm n}\hat{\boldsymbol{\Phi}}_{\rm y}\left(k,l-1\right) + \mathbf{y}\left(k,l\right)\mathbf{y}^{H}\left(k,l\right)\label{eq:SOS_RnUpdate}\,.
\end{align}
Speech-plus-noise and noise-only TF bins are distinguished based on the speech presence probabilities \cite{Gerkmann2012} in the head-mounted microphones, which are averaged and thresholded per TF bin. For the stationary speaker scenario initialization effects are mitigated by using the first half of the signal as initialization period and evaluating the performance on the second half only. The prototype head-mounted RTF vectors $\bar{\mathbf{g}}_{\rm h}\left(k,\theta_{i}\right)$ were generated using the database of binaural anechoic room impulse responses in \cite{Kayser2009} with an angular resolution of $\ang{5}$ ($I=72$).

As performance measure we use the localization accuracy, i.e., the percentage of correctly localized frames, defined as
\begin{equation}
	\label{eq:defAcc}
	\text{\rm ACC} = \frac{1}{L}\sum_{l=1}^{L}U\left(\Delta\theta-f\left(\lvert\hat{\theta}_{\rm s}\left(l\right)-\theta_{\rm s}\left(l\right)\rvert\right)\right)\times 100\%\,,
\end{equation}
where $U$ is the Heaviside step function and $f\left(\cdot\right)$ is a circular wrapping function to ensure an absolute error smaller than $\ang{180}$. As tolerance we used $\Delta\theta = \ang{5}$, which corresponds to the resolution of the prototype RTF vectors.

\subsection{DOA estimation accuracy}
\label{sec:exp_res}
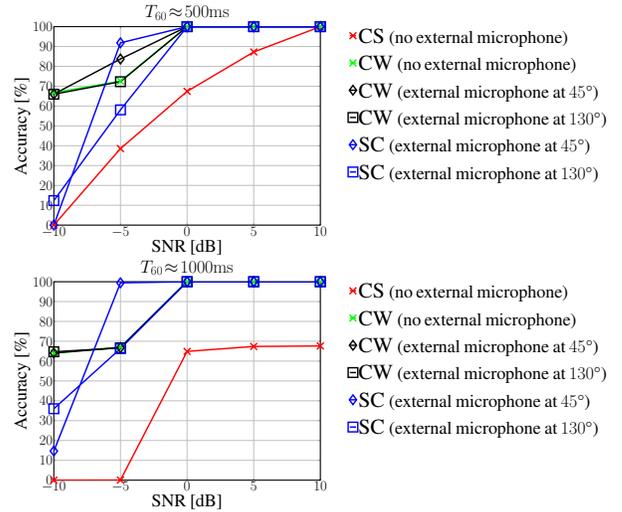
\begin{figure}[t]
	\centering
%
%
%
\begin{tikzpicture}[scale=0.135]

\begin{axis}[%
width=10.333in,
height=7.7in,
at={(1.733in,1.039in)},
scale only axis,
xmin=-10,
xmax=10,
xtick={-10,  -5,   0,   5,  10},
xlabel style={font=\color{white!15!black}, font = \fontsize{50}{1}\selectfont},
xticklabel style={font = \fontsize{40}{1}\selectfont},
xlabel={SNR [dB]},
ymin=0,
ymax=100,
ytick={  0,  10,  20,  30,  40,  50,  60,  70,  80,  90, 100},
ylabel style={font=\color{white!15!black}, font = \fontsize{50}{1}\selectfont},
ylabel={Accuracy [\%]},
yticklabel style={font = \fontsize{40}{1}\selectfont},
axis background/.style={fill=white},
title style={font = \fontsize{50}{1}\selectfont,yshift=2ex},
title={$T_{\rm 60}\approx\SI{500}{\milli\second}$},
xmajorgrids,
ymajorgrids,
legend style={at={(1.1,1)},anchor=north west, legend cell align=left, align=left, draw=none, font = \fontsize{60}{1}\selectfont,row sep=30pt}
]
\addplot [color=red, line width=4.0pt, mark=x, mark options={solid, red}, mark size=13.5pt]
  table[row sep=crcr]{%
-10	0\\
-5	38.5514777525171\\
0	67.3920103929847\\
5	87.2361156219552\\
10	100\\
};
\addlegendentry{CS \fontsize{50}{1}\selectfont(no external microphone)}

\addplot [color=green, line width=4.0pt, mark=x, mark options={solid, green}, mark size=13.5pt]
  table[row sep=crcr]{%
-10	66.7424488470283\\
-5	72.5235466060409\\
0	100\\
5	100\\
10	100\\
};
\addlegendentry{CW \fontsize{50}{1}\selectfont(no external microphone)}

\addplot [color=black, line width=4.0pt, mark=diamond, mark options={solid, black}, mark size=13.5pt]
  table[row sep=crcr]{%
-10	65.9954530691783\\
-5	83.6960051964924\\
0	100\\
5	100\\
10	100\\
};
\addlegendentry{CW \fontsize{50}{1}\selectfont(external microphone at $\ang{45}$)}

\addplot [color=black, line width=4.0pt, mark=square, mark options={solid, black}, mark size=13.5pt]
  table[row sep=crcr]{%
-10	65.9629749918805\\
-5	72.2637219876583\\
0	100\\
5	100\\
10	100\\
};
\addlegendentry{CW \fontsize{50}{1}\selectfont(external microphone at $\ang{130}$)}

\addplot [color=blue, line width=4.0pt, mark=diamond, mark options={solid, blue}, mark size=13.5pt]
  table[row sep=crcr]{%
-10	0\\
-5	91.8480025982462\\
0	100\\
5	100\\
10	100\\
};
\addlegendentry{SC \fontsize{50}{1}\selectfont(external microphone at $\ang{45}$)}

\addplot [color=blue, line width=4.0pt, mark=square, mark options={solid, blue}, mark size=13.5pt]
  table[row sep=crcr]{%
-10	12.3416693731731\\
-5	58.0708022085093\\
0	100\\
5	100\\
10	100\\
};
\addlegendentry{SC \fontsize{50}{1}\selectfont(external microphone at $\ang{130}$)}

\end{axis}

\end{tikzpicture}%
%
	\vskip0.05cm
	\centering
%
%
\definecolor{mycolor1}{rgb}{0.50000,0.80000,0.10000}%
\definecolor{mycolor2}{rgb}{0.10000,0.30000,0.80000}%
\begin{tikzpicture}[scale=0.135]

\begin{axis}[%
width=10.333in,
height=7.7in,
at={(1.733in,1.039in)},
scale only axis,
xmin=-10,
xmax=10,
xtick={-10,  -5,   0,   5,  10},
xlabel style={font=\color{white!15!black}, font = \fontsize{50}{1}\selectfont},
xticklabel style={font = \fontsize{40}{1}\selectfont},
xlabel={SNR [dB]},
ymin=0,
ymax=100,
ytick={  0,  10,  20,  30,  40,  50,  60,  70,  80,  90, 100},
ylabel style={font=\color{white!15!black}, font = \fontsize{50}{1}\selectfont},
ylabel={Accuracy [\%]},
yticklabel style={font = \fontsize{40}{1}\selectfont},
axis background/.style={fill=white},
title style={font = \fontsize{50}{1}\selectfont,yshift=2ex},
title={$T_{\rm 60}\approx\SI{1000}{\milli\second}$},
xmajorgrids,
ymajorgrids,
legend style={at={(1.1,1)},anchor=north west, legend cell align=left, align=left, draw=none, font = \fontsize{60}{1}\selectfont,row sep=30pt}
]
\addplot [color=red, line width=4.0pt, mark=x, mark options={solid, red}, mark size=13.5pt]
  table[row sep=crcr]{%
-10	0\\
-5	0.061652281134402\\
0	64.8890258939581\\
5	67.3859432799014\\
10	67.632552404439\\
};
\addlegendentry{CS \fontsize{50}{1}\selectfont(no external microphone)}

\addplot [color=green, line width=4.0pt, mark=x, mark options={solid, green}, mark size=13.5pt]
  table[row sep=crcr]{%
-10	63.9950678175092\\
-5	67.0776818742293\\
0	100\\
5	100\\
10	100\\
};
\addlegendentry{CW \fontsize{50}{1}\selectfont(no external microphone)}

\addplot [color=black, line width=4.0pt, mark=diamond, mark options={solid, black}, mark size=13.5pt]
  table[row sep=crcr]{%
-10	64.0567200986437\\
-5	66.8002466091245\\
0	100\\
5	100\\
10	100\\
};
\addlegendentry{CW \fontsize{50}{1}\selectfont(external microphone at $\ang{45}$)}

\addplot [color=black, line width=4.0pt, mark=square, mark options={solid, black}, mark size=13.5pt]
  table[row sep=crcr]{%
-10	64.7040690505549\\
-5	66.6769420468557\\
0	100\\
5	100\\
10	100\\
};
\addlegendentry{CW \fontsize{50}{1}\selectfont(external microphone at $\ang{130}$)}

\addplot [color=blue, line width=4.0pt, mark=diamond, mark options={solid, blue}, mark size=13.5pt]
  table[row sep=crcr]{%
-10	14.6732429099877\\
-5	99.4451294697904\\
0	100\\
5	100\\
10	100\\
};
\addlegendentry{SC \fontsize{50}{1}\selectfont(external microphone at $\ang{45}$)}

\addplot [color=blue, line width=4.0pt, mark=square, mark options={solid, blue}, mark size=13.5pt]
  table[row sep=crcr]{%
-10	36.0049321824907\\
-5	66.3070283600493\\
0	100\\
5	100\\
10	100\\
};
\addlegendentry{SC \fontsize{50}{1}\selectfont(external microphone at $\ang{130}$)}

\end{axis}

\end{tikzpicture}%
%
	\vspace{-3mm}
	\caption{Average localization accuracy for all considered RTF vector estimation methods for different SNRs. Top: $T_{\rm 60}\approx\SI{500}{\milli\second}$, bottom: $T_{\rm 60}\approx\SI{1000}{\milli\second}$.}
	\vskip-0.34cm
	\label{fig:resStat}
\end{figure}
For the stationary speaker scenario, Fig. \ref{fig:resStat} depicts the lo\-calization accuracy (averaged over the three speaker positions) for all considered RTF vector estimation methods as a function of SNR for two reverberation times $\left(T_{\rm 60}\approx\SI{500}{\milli\second},~T_{\rm 60}\approx\SI{1000}{\milli\second}\right)$. For the CW and SC methods exploiting the external microphone, the performance is shown for both considered positions of the external microphone $\left(\ang{45},\ang{130}\right)$. First, it can be observed that for both reverberation times and for all SNRs the CW and SC methods outperform the CS method. Second, it can be observed that for both reverberation times and for all SNRs except $-10 \: \si{dB}$ the SC method yields a similar localization accuracy as the CW methods. The performance of the SC method appears to depend more on the position of the external microphone than the performance of the CW method, which is especially noticeable at $\text{SNR}= -5\:\si{dB}$.

\begin{figure}[t]
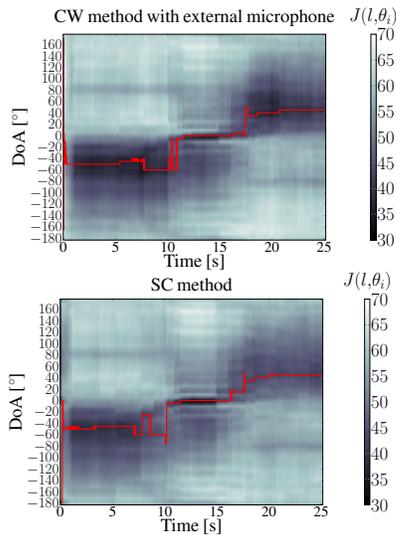

	\centering
	\input{Results_moving_CW_eMic.tex}
	\input{Results_moving_SC.tex}
	\vspace{-3mm}
	\caption{Localization performance for the moving speaker scenario for $\text{SNR} = 0 \: \si{dB}$ and $T_{\rm 60}\approx\SI{400}{\milli\second}$. Top: CW method with external microphone, bottom: SC method}
	\vskip-0.34cm
	\label{fig:resMov}
\end{figure}

For the moving speaker scenario, we only consider the SC and CW methods incorporating the external microphone. Fig. \ref{fig:resMov} depicts for an SNR of $0 \: \si{dB}$ and $T_{\rm 60}\approx\SI{400}{\milli\second}$ the time-varying estimated DOA $\hat{\theta}_{\rm s}\left(l\right)$ (solid red line), while the gray background encodes the cost function $J\left(l,\theta_{i}\right)$ in \eqref{eq:doaEst}. Although no exact ground-truth DOA is available for the moving speaker scenario, it can be observed that the moving speaker can be localized well using both considered RTF vector estimation methods. In addition, it can be observed that a higher localization confidence is obtained using the SC method than using the CW method, because the region of small Hermitian angles around the estimated DOA is more confined for the SC method than for the CW method.

The DOA estimation results for the stationary and moving speaker scenario show that the low-complexity SC method yields a comparable performance as the CW method, which is in line with the beamforming results reported in \cite{Goessling2018,Goessling2019}.

\section{Conclusions}
\label{seq:sum}
In this paper we analyzed the DOA estimation performance based on several RTF vector estimation methods for a binaural hearing aid setup with an external microphone that is not restricted to be close to the target speaker. More in particular, we compared the performance of the state-of-the-art CW and CS methods with the SC method. To estimate the DOA from the estimated head-mounted RTF vector, we proposed to minimize the frequency-averaged Hermitian angle between the estimated head-mounted RTF vector and anechoic prototype head-mounted RTF vectors for several directions. Experimental results with real-world data for stationary and moving speaker scenarios show that exploiting the external microphone using the SC method yields a similar DOA estimation accuracy as the CW method at a lower computational complexity.


\begin{thebibliography}{00}
\bibitem{Marquardt2016}D. Marquardt and S. Doclo, "Performance comparison of bilateral and binaural MVDR-based noise reduction algorithms in the presence of DOA estimation errors," \textit{Proc. ITG Symposium on Speech Communication}, Paderborn, Germany, Oct. 2016, pp. 1-5.	
%
\bibitem{Raspaud2010}M. Raspaud, H. Viste and G. Evangelista, "Binaural source localization by joint estimation of ILD and ITD," \textit{IEEE Trans. Audio, Speech, and Language Processing}, vol. 18, no. 1, pp. 68-77, Jan. 2010.
%
\bibitem{May2012}T. May, S. van de Par and A. Kohlrausch, "A binaural scene analyzer for joint localization and recognition of speakers in the presence of interfering noise sources and reverberation," \textit{IEEE Trans. Audio, Speech, and Language Processing}, vol. 20, no. 7, pp. 2016-2030, Sept. 2012.
%
\bibitem{Kayser2014}H. Kayser and J. Anemüller, "A discriminative learning approach to probabilistic acoustic source localization," \textit{Proc. International Workshop on Acoustic Signal Enhancement} (\textit{IWAENC}), Juan-les-Pins, France, Sept. 2014, pp. 99-103.
%
\bibitem{Marquardt2017}D. Marquardt and S. Doclo, "Noise power spectral density estimation for binaural noise reduction exploiting direction of arrival estimates," \textit{Proc. IEEE Workshop on Applications of Signal Processing to Audio and Acoustics} (\textit{WASPAA}), New Paltz, USA, Oct. 2017, pp. 234-238.
%
\bibitem{Ma2018}N. Ma, J. A. Gonzalez and G. J. Brown, "Robust binaural localization of a target sound source by combining spectral source models and deep neural networks," \textit{IEEE/ACM Trans. Audio, Speech, and Language Processing}, vol. 26, no. 11, pp. 2122-2131, Nov. 2018.
%
\bibitem{Varzandeh2020}R. Varzandeh, K. Adiloğlu, S. Doclo and V. Hohmann, "Exploiting periodicity features for joint detection and DOA estimation of  speech sources using convolutional neural networks," \textit{Proc. IEEE International Conference on Acoustics, Speech and Signal Processing} (\textit{ICASSP}), Barcelona, Spain, May 2020, pp. 566-570.
%
\bibitem{Braun2015}S. Braun, W. Zhou and E. A. P. Habets, "Narrowband direction-of-arrival estimation for binaural hearing aids using relative transfer functions," \textit{Proc. IEEE Workshop on Applications of Signal Processing to Audio and Acoustics} (\textit{WASPAA}), New Paltz, USA, Oct. 2015, pp. 1-5.
%
\bibitem{Farmani2018}M. Farmani, M. S. Pedersen, Z. Tan and J. Jensen, "Bias-compensated informed sound source localization using relative transfer functions," \textit{IEEE/ACM Trans. Audio, Speech, and Language Processing}, vol. 26, no. 7, pp. 1275-1289, July 2018.
%
\bibitem{Cohen2004}I. Cohen, "Relative transfer function identification using speech signals," \textit{IEEE Trans. Speech and Audio Processing}, vol. 12, no. 5, pp. 451-459, Sept. 2004.

\bibitem{Markovich2009}S. Markovich, S. Gannot and I. Cohen, "Multichannel eigenspace beamforming in a reverberant noisy environment with multiple interfering speech signals," \textit{IEEE Trans. Audio, Speech, and Language Processing}, vol. 17, no. 6, pp. 1071-1086, Aug. 2009.
%
\bibitem{Krueger2011}A. Krueger, E. Warsitz and R. Haeb-Umbach, "Speech enhancement with a GSC-like structure employing eigenvector-based transfer function ratios estimation," \textit{IEEE Trans. Audio, Speech, and Language Processing}, vol. 19, no. 1, pp. 206-219, Jan. 2011.
%
\bibitem{Varzandeh2017}R. Varzandeh, M. Taseska and E. A. P. Habets, "An iterative multi\-channel subspace-based covariance subtraction method for relative transfer function estimation," \textit{Proc. Joint Workshop on Hands-free Speech Communication and Microphone Arrays} (\textit{HSCMA}), San Francisco, USA, Mar., 2017, pp. 11-15.
%
\bibitem{Golan2018}S. Markovich-Golan, S. Gannot and W. Kellermann, "Performance analysis of the covariance-whitening and the covariance-subtraction methods for estimating the relative transfer function," \textit{Proc. European Signal Processing Conference} (\textit{EUSIPCO}), Rome, Italy, Sept. 2018, pp. 2499-2503.
%
\bibitem{Goessling2018}N. Gößling and S. Doclo, “Relative transfer function estimation exploiting spatially separated microphones in a diffuse noise field,” \textit{Proc. International Workshop on Acoustic Signal Enhancement} (\textit{IWAENC}), Tokyo, Japan, Sept. 2018, pp. 146–150.
%
\bibitem{Goessling2020}N. Gößling, "Binaural beamforming algorithms and parameter estimation methods exploiting external microphones", PhD Thesis, University of Oldenburg, Germany, Oct. 2020. 
%
\bibitem{Li2016}X. Li, L. Girin, R. Horaud and S. Gannot, "Estimation of the direct-path relative transfer function for supervised sound-source localization," \textit{IEEE/ACM Trans. Audio, Speech, and Language Processing}, vol. 24, no. 11, pp. 2171-2186, Nov. 2016. 
%
\bibitem{Hammer2020}H. Hammer, S. E. Chazan, J. Goldberger, and S. Gannot, “FCN approach for dynamically locating multiple speakers,” Aug. 2020, [Online], available: https://arxiv.org/abs/2008.11845.
%
\bibitem{Avargel2007}Y. Avargel and I. Cohen, "On multiplicative transfer function approxi\-mation in the short-time Fourier transform domain," \textit{IEEE Signal Processing Letters}, vol. 14, no. 5, pp. 337-340, May 2007.
%
\bibitem{Duda1998}R. O. Duda and W. L. Martens, “Range dependence of the response of a spherical head model,” \textit{Journal of the Acoustical Society of America}, vol. 104, no. 5, pp. 3048–3058, Nov. 1998.
%
\bibitem{Gerkmann2012}T. Gerkmann and R. C. Hendriks, “Unbiased MMSE-based noise power estimation with low complexity and low tracking delay,” \textit{IEEE Trans. Audio, Speech, and Language Processing}, vol. 20, no. 4, pp. 1383–1393, May 2012.
%
\bibitem{Kayser2009}H. Kayser, S. D. Ewert, J. Anemüller, T. Rohdenburg, V. Hohmann, and B. Kollmeier, “Database of multichannel In-Ear and Behind-the-Ear head-related and binaural room impulse responses,” \textit{EURASIP Journal on Advances in Signal Processing}, vol. 2009, pp. 1–10, Jan. 2009.
%
\bibitem{Goessling2019} N. Gößling and S. Doclo, "RTF-steered binaural MVDR beamforming incorporating an external microphone for dynamic acoustic scenarios," \textit{Proc. IEEE
International Conference on Acoustics, Speech, and Signal Processing} (\textit{ICASSP}),
Brighton, UK, May 2019, pp. 416–420.
\end{thebibliography}
\end{document}